# Demonstration of an iron-pnictide bulk superconducting magnet capable of trapping over 1 T

J. D. Weiss[1*], A. Yamamoto[2,3,4**], A. A. Polyanskii[1], R. B. Richardson[1], D. C. Larbalestier[1], and E. E. Hellstrom[1]

[1]Applied Superconductivity Center, National High Magnetic Field Laboratory, Florida State University, Tallahassee, FL, 32310, USA

(*) E-mail: Weiss@magnet.fsu.edu

[2]Department of Applied Chemistry, The University of Tokyo, 7-3-1 Hongo, Bunkyo, Tokyo, 113-8656, Japan

[3] JST-PRESTO, 4-1-8 Honcho, Kawaguchi, Saitama, 332-0012, Japan

[4]Department of Applied Physics, Tokyo University of Agriculture and Technology,

2-24-16 Nakacho, Koganei, Tokyo 184-8588, Japan

(**) E-mail: Akiyasu@cc.tuat.ac.jp

**Abstract.** A trapped field of over 1 T at 5 K and 0.5 T at 20 K has been measured between a stack of magnetized cylinders of bulk polycrystalline $Ba_{0.6}K_{0.4}Fe_2As_2$ superconductors 10 mm in diameter and 18 mm in combined thickness. The trapped field showed a low magnetic creep rate (~3% after 24 hours at 5 K), while magneto-optical imaging revealed a trapped field distribution corresponding to uniform macroscopic current loops circulating through the sample. The superconductors were manufactured by hot isostatic pressing of pre-reacted powders using the scalable powder-in-tube technique. A high Vickers hardness of ~3.5 GPa and a reasonable fracture toughness of ~2.35 MPa m$^{0.5}$ were measured. Given the untextured polycrystalline nature of the cylinders and their large irreversibility field (> 90 T), it is expected that larger bulks could trap fields in excess of 10 T.

Demonstration of an iron-pnictide bulk superconducting magnet capable of trapping over 1 T

**Introduction**

Since their discovery in 2008,[1] a tremendous research effort has been made to synthesize and study Iron-based superconductors. Much of this effort has been driven by reports of properties that are very appealing for applications including low anisotropy around 1-2,[2] high upper critical fields ($H_{c2}$) in excess of 90 T,[3,4] and intrinsic critical current densities above 1 MAcm$^{-2}$ (0 T, 4.2 K).[5] Unfortunately, soon after their discovery, grain boundaries were observed to block current, similar to the behavior of rare-earth barium cuprate (REBCO) materials like $YBa_2Cu_3O_{7-x}$ (YBCO), but to a somewhat lesser extent.[6–8] Remarkably, fine-grain, randomly oriented K-doped $BaFe_2As_2$ (Ba122) has been synthesized with a global critical current density around 10 kAcm$^{-2}$ (4.2 K, 10 T)[9,10] while textured tapes of K-doped Ba122 and $SrFe_2As_2$ (Sr122) have now been produced that raise $J_c$ by another order of magnitude.[11,12] The untextured $J_c$ values are within an order of magnitude of those needed for applications as highly desirable round wires. The better $J_c$ values of tapes suggest some grain boundary improvement, perhaps due to texturing produced by the rolling and/or pressing processes. However, recent atom probe field ion microscopy study of our carefully made untextured bulks shows significant O impurity segregation to grain boundaries and As and Fe perturbations within several nm of the GB plane.[13] Thus there is evidence for both intrinsic and extrinsic degradation of superconductivity at the GB in the Fe-based superconductors.

Although the major route to utilizing superconductors for generating fields is by using wire-wound magnets, another way is to produce bulk samples, generally as disks, that can be used as trapped field magnets. Iron-based permanent magnets are limited by their ferromagnetic saturation magnetization, and are therefore not capable of producing fields much greater than 1 T. However, persistent currents induced by external fields can be trapped inside a superconductor to produce magnetic fields ($B^{trapped}$) that scale with the size of the current loops flowing in the bulk:

$$B^{trapped} = A\mu_0 J_c^{bulk} r$$

where A is a geometrical factor, $\mu_0$ is the permeability of vacuum, $J_c^{bulk}$ is the bulk or globally circulating critical current density, and r is the radius of the sample. High field bulk magnets thus require high $J_c^{bulk}$ and/or large r with a well-defined geometry. The field trapping ability is then limited mostly by the ability of the material to develop high $J_c(H)$ at high fields and since Ba122 has $H_{c2}$ > 90 T,[4] $T_c$ of 38 K and reasonable mechanical strength (the electromagnetic Lorentz force results in stresses proportional to the local values of $J_c$, B, and r), we wished to explore the potential of bulk Ba122 magnets by synthesizing

Demonstration of an iron-pnictide bulk superconducting magnet capable of trapping over 1 T

and characterizing cylindrical bulks. Here we report the first Iron-pnictide superconducting magnet capable of trapping over 1 T at 5 K. Electromagnetic and mechanical measurements suggest that the material properties are suitable for making larger bulk magnets that can be magnetized to trap very strong magnetic fields (> 10 T).

**Experimental details**

Ba, K, Fe, and As were combined in a molar ratio of 0.6:0.42:2:2 and reacted together by a mechanochemical reaction followed by sintering in a hot isostatic press (HIP) at 600 °C using procedures described elsewhere.[10] After bulk synthesis and subsequent re-milling, approximately 3-5 g of Ba122 powder was pressed into 15.9 mm diameter pellets and then further densified in a cold isostatic press (CIP) at 276 MPa. These were then wrapped with Ag foil and inserted into a steel tube that was carefully machined to the diameter of the pellet + foil. Both ends of the steel tube were welded shut under vacuum using chamfered plugs that helped the steel tube to compress the pellet. After welding, the tubes were swaged and CIPped to further shape and densify them, reducing the diameter of the samples ~10%. Finally, the samples were sintered for 10 hours at 600 °C in the HIP. After heat treatment, the steel tubes were sliced with a diamond saw to reveal the pellet surfaces on which their Vickers hardness (HV) was measured using loads from 25 g to 2000 g. Light and scanning electron microscopy was used to study and measure the micro-indentations.

Magneto optical imaging was used to image the uniformity of the local field profile produced by magnetization currents induced by field-cooling into the superconducting state in a field of 120 mT applied perpendicular to the sample surface and then removing the magnetic field. Due to the limited size of the cryostat, magneto optical imaging was done on a 3.7 mm thick sample. Then, disk-shaped Ba122 bulks ~10 mm in diameter and ~18.4 mm in total thickness were vertically stacked on either side of a spacer containing a cryogenic Hall sensor to measure the perpendicular magnetic flux density between the pellets. Another Hall sensor was mounted on the outside end of the stack. The stack was cooled to ~5 K by a GM cryocooler under an external field ($H_{app}$) of 8 T, and the external field was subsequently removed. After reduction of the external field to zero, the magnetic flux density trapped in the bulk was measured at the center of the spacer as a function of increasing temperature (0.2 K/min) and separately as a function of time at constant temperature. For the magnetic hysteresis loop measurement, the sample was

Demonstration of an iron-pnictide bulk superconducting magnet capable of trapping over 1 T

zero-field cooled to 5 K and the flux density at the center of the sample stack recorded as a function of increasing and decreasing external field.

**Results**

To assess the mechanical properties of the Ba122 bulks, room temperature Vickers Hardness tests were performed. The average HV was 3.5 (± 0.2) GPa. Cracks were observed propagating from the corners of the indentations as is typical in Vickers Hardness testing of brittle materials. Figure 1a shows a light micrograph of one of the bulks with thickness = 3.7 mm and figure 1b and 1c show magneto optical images after cooling in 120 mT perpendicular field to 11 and to 20 K and then reducing the external field to zero. The images are featureless in the center of each image and with a uniform gradient at each edge, thus showing rather uniform flux gradients across the disk.

Figure 2 shows the trapped field (at $H_{app}$ = 0) measured by Hall sensors placed on the bottom surface (H1) and between (H2) the stack of Ba122 bulk magnets after field cooling ($H_{app}$ = 8 T) to 5K, reducing the magnetizing field to zero and then increasing the temperature at 0.2 K/min. It should be noted that H2 was placed 3.7 mm away from H1, which is 5.5 mm from the center of the stack where the maximum field would be expected. At 5 K, the bulk stack trapped 0.68 T at the center of the outer surface (H1) and 1.02 T on the cylinder axis between the bulks (H2). The trapped field decreased with increasing temperature and vanished at $T_c$ ~33 K. The average macroscopic current density at 5 K was estimated by a Biot-Savart approximation[14] using the total thickness of the magnet stack and the experimentally obtained trapped field of H1 and found to be ~ 50 kAcm$^{-2}$. This matches $J_c$ obtained by local magnetization measurements made on small bulk samples at T = 4.2 K and $H_{app}$ = 0.6 T.[10]

On field cooling, no flux jumps were observed at a ramp rate of -1.8 T/h. In one instance, an unexpected quench of the magnetizing magnet resulted in a rapid collapse of the external field from 1.5 T to 0 T, corresponding to a ramp rate of > 2 T/sec. Despite the sudden removal of flux, the trapped field value quickly shifted to the expected critical state and the subsequent trapped field creep behavior was identical to data taken during the controlled process where $H_{app}$ was slowly removed.

Figure 3 shows the magnetic flux density inside the stack of the initially zero-field cooled bulk samples as a function of increasing and decreasing external field. The samples showed strong, quasi-diamagnetic shielding below $H_{app}$ = 0.7 T and the hysteresis loop remained open to $H_{app}$ > 8 T (see inset),

Demonstration of an iron-pnictide bulk superconducting magnet capable of trapping over 1 T

the maximum field of our magnet. Figure 4a shows that the trapped field decayed approximately 3% after one day at 5 K. Figure 4b shows the normalized relaxation of trapped field as a function of time at 5 K, 10 K, and 20 K. The magnet creep rate increased with increased temperature due to thermally activated flux depinning. The trapped field decayed ~7 % after 1 day at 20 K.

**Discussion**

Use of bulk superconductors to generate significant magnetic fields has lagged well behind the use of coils wound with wires where the magnetic field is largely determined by the transport current in the winding. However, the generation of trapped fields exceeding 15 T in single crystal REBCO bulks with diameters of order 25 mm demonstrates their significant potential in applications such as bearings where driven magnets would not be easily applicable. Moreover such fields are generated at temperatures exceeding 20 K where no low temperature superconductor is still superconducting. The well-known weak link problem of the cuprates restricts such magnets to single crystals and in this respect the recent work on making larger diameter (> 30 mm) polycrystalline samples of $MgB_2$ that can now trap more than 3 T is very promising too.[15–17] The purpose of this paper is to indicate that the Fe-based superconductor, $(Ba_{0.6}K_{0.4})Fe_2As_2$, can also generate tesla-scale fields with a polycrystalline microstructure. Because its $T_c$ is the same as $MgB_2$ and because its $H_{c2}$ is many times higher (>90 T versus <30 T), it may offer major advantages for large diameter magnets that cannot be supplied by either REBCO or $MgB_2$.

Our bulk Ba122 magnets were processed by a scalable, versatile, and low-cost technique using ball milling, CIPping, and HIPping, all common industrial ceramic processing techniques. The samples were all polycrystalline, thus avoiding any of the complex processing needed for single crystals. Use of the powder-in-tube technique and low-temperature reaction allow for several bulks to be produced in a single batch from which disks can then be sliced to a desired thickness. Clearly, the limit to the length and diameter of the disks is set by the HIP size and since some HIPs can reach >1 m in dimension, we believe that this technique has significant scale up potential. In addition, the use of a metal sheath adds a reinforcing ring that can easily be designed to further improve the mechanical strength of the bulk, as has proven invaluable for trapping high fields using REBCO bulks.[18] High strength bulks and external reinforcement are important because the large currents that produce the trapped fields produce stresses of order $J_c$ x B. Due to their brittle failure mechanics, superconductors with high fracture toughness ($K_C$) are desired for high field applications. A $K_C$ of ~2.35 (± 0.14) MPa m$^{0.5}$ was calculated from the length of micro-cracks propagating from the corners of the micro-indentations according to the following formula:[19]

Demonstration of an iron-pnictide bulk superconducting magnet capable of trapping over 1 T

$$K_C = 0.0726 \frac{P}{C^{3/2}}$$

where P is loading force in N, C is the distance between center of indention and the tip of the crack in m, and 0.0726 is a calibration constant. This fracture toughness exceeds single crystal Mn-doped Ba122,[20] HIPped MgB$_2$,[21] bulk top-seeded melt-grown YBCO,[22] and is about equal to polycrystalline Al$_2$O$_3$.[23] There thus seems to be no mechanical reasons for ruling out the use of K-doped Ba122 for trapped field applications.

A key issue is the uniformity of current flow in the bulk. Extensive study of Ba122 bulks over several years has shown us that there is some obstruction to the flow of currents on a fine scale, as judged by measurements of hysteresis in the field-increasing/field-decreasing $I_c(H)$ characteristics of wires. However, magneto-optical examinations have shown that current flow is uniform on the scales of several μm to several mm observable by magneto-optics.[9,10] The smaller scale is about an order of magnitude greater than the 100-300 nm grain size of our Ba122 material. The flux distributions observed by magneto optical imaging in figure 1 reinforce this conclusion because they show that the trapped field is distributed rather uniformly around the entire bulk sample. Moreover, no flux avalanches were observed, even when the external magnet quenched. It may be that Ba122 is less susceptible to flux jumps and avalanches than MgB$_2$ and YBCO bulks but conclusive evidence on this point will require larger samples and higher trapped fields. One property that requires further study is the decay rate of the trapped field, which is higher (~3% after a day at 5 K) than in MgB$_2$ (~1.5% at 20 K).[15]

Returning therefore to consideration of where Ba122 could fit as a new trapped field magnet material, it is worth considering various scenarios for the trapped field. Figure 5 (a) shows a comparison of $J_c(H)$ data taken from the literature.[10,24,25] Figure 5 (b) shows a calculation of maximum $B^{trapped}$ as a function of r for K-doped Ba122 and MgB$_2$ using the bulk $J_c(H)$ data presented in figure 5(a). This calculation is for an infinitely long cylinder and takes a radial field dependence of $J_c(H)$ into account. In the infinite long cylinder geometry, local current density j(x) varies with respect to radial direction x (0 < x < r) due to self-field but does not change with respect to the circumferential and length directions. Thus, the local flux density b(x) and j(x) can be calculated as:

$$b(x) = \int \mu_0 j(x) dx \, , j(x) = J_c(b(x))$$

The maximum $B^{trapped}$ at the center of a bulk is given as b(r). To allow a full evaluation of MgB$_2$, we included $J_c$ data for C-doped MgB$_2$ bulks, though MgB$_2$ bulk magnets in the literature are typically

Demonstration of an iron-pnictide bulk superconducting magnet capable of trapping over 1 T

undoped and therefore have not demonstrated the ability to trap the high fields suggested by figure 5 (b) at 4.2 K. So far the maximum reported trapped fields at 20K are a little over 3 T.[15–17] Currently, mechanically reinforced REBCO materials have produced the highest fields. Their record fields (> 17 T)[18,26] occur at temperatures > 20 K, but are limited in size (r ≤ 50 mm) because single crystals are needed to avoid drastic loss of the $J_c$ by the presence of any grain boundaries. In contrast, $MgB_2$ is not subject to intrinsic GB current blocking and so it can be manufactured as large diameter polycrystalline bulks.[14–17] However, the maximum trapped field calculated using the above equation and the data from figure 5 (a) is barely above 6 T at 20 K because the low $H_{c2}$ strongly limits $J_c(H)$. Ba122 in principle offers significant advantages over $MgB_2$ because of its much higher $H_{c2}$ of about 80 T at 20 K.[4] While $MgB_2$ outperforms Ba122 at low fields and small r, Ba122 bulks with r ≥ 100 mm would be capable of trapping higher fields than $MgB_2$, even at 20 K. Such large bulks would be useful for magnetic levitation, in energy storage applications, and could provide high fields in compact magnetic resonance devices.

**Conclusions**

In summary, we successfully synthesized the first bulk iron-pnictide demonstration magnet capable of trapping over 1 T (5 K) and 0.5 T (20K) with fine-grain polycrystalline $Ba_{0.6}K_{0.4}Fe_2As_2$ by using a scalable technique that could generate much larger samples. Magneto optical imaging showed rather uniform macroscopic currents circulating throughout the entire sample. The time dependence of the trapped field showed a low magnetic creep rate (~3% after 24 hours at 5 K). Vickers hardness indentations indicate a hardness ~3.5 GPa and a fracture toughness ~2.35 MPa m$^{0.5}$. Larger bulks are expected to trap even higher fields, given the high $H_{c2}$ of K-doped $BaFe_2As_2$. Modest improvements to $J_c(H)$ will make Ba122 bulks very competitive against REBCO and $MgB_2$ for bulk magnet applications.

**Acknowledgments**

We would like to thank W. L. Starch and B. H. Hainsey for technical. The work at the National High Magnetic Field Laboratory was supported by NSF (No. DMR-1306785) to E. E. Hellstrom and the facilities of the NHMFL are supported by the State of Florida and by the NSF through a facility grant (No. DMR-1157490). The work at the University of Tokyo was carried out under the JST-PRESTO project and partially supported by JSPS.

Demonstration of an iron-pnictide bulk superconducting magnet capable of trapping over 1 T

Demonstration of an iron-pnictide bulk superconducting magnet capable of trapping over 1 T

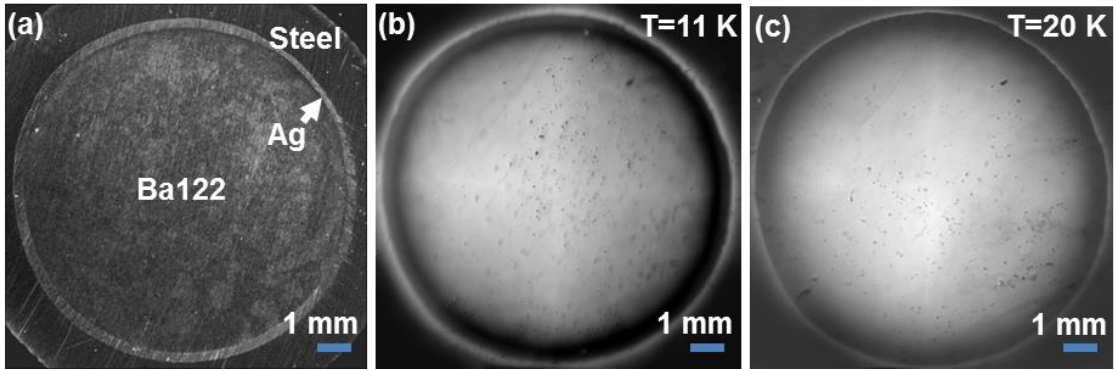

Figure 1 - (a) Light microscopy image of a polished surface of the disk-shaped K-doped Ba122 bulk sample (10 mm diameter and 3.7 mm thick). (b), (c) magneto optical images at (b) 11 K and (c) 20 K after field cooling under 120 mT and then reducing the external field to zero. The images show macroscopically uniform field in the center of each disk corresponding to the trapped field of 120 mT and uniform field gradients at the perimeter.

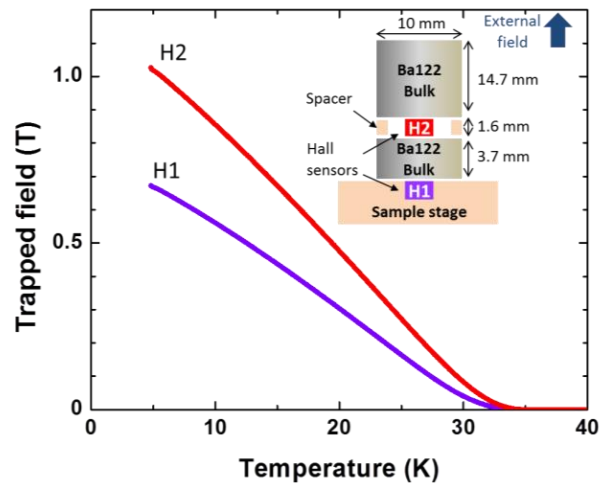

Figure 2 - Trapped field as a function of increasing temperature for the bulk sample stack that was field-cooled to 5 K in 8 T. After reducing the external field to zero, the sample stack was heated at 0.2 K/min and the relaxation of the trapped field was measured. A schematic of the sample and Hall probe arrangement is shown.

Demonstration of an iron-pnictide bulk superconducting magnet capable of trapping over 1 T

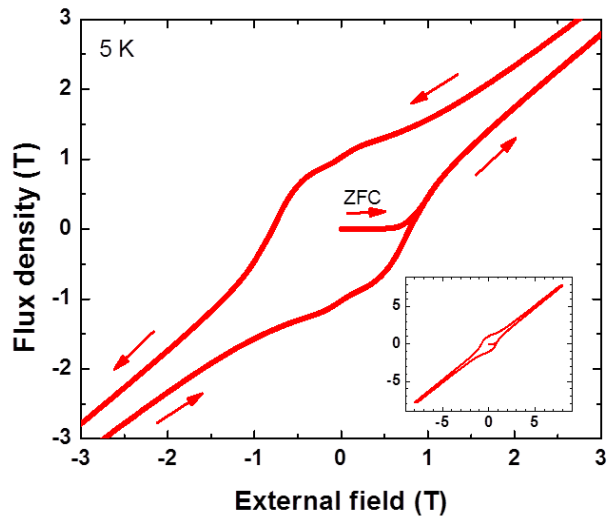

Figure 3 - Trapped field hysteresis loop obtained at 5 K. The sample was zero-field cooled to 5 K and the flux density inside the sample stack measured by H2 was recorded as a function of increasing and decreasing external field. The inset shows that the hysteresis loop remains open up to our maximum applied field of 8 T, consistent with the very high critical fields measured in bulk samples.

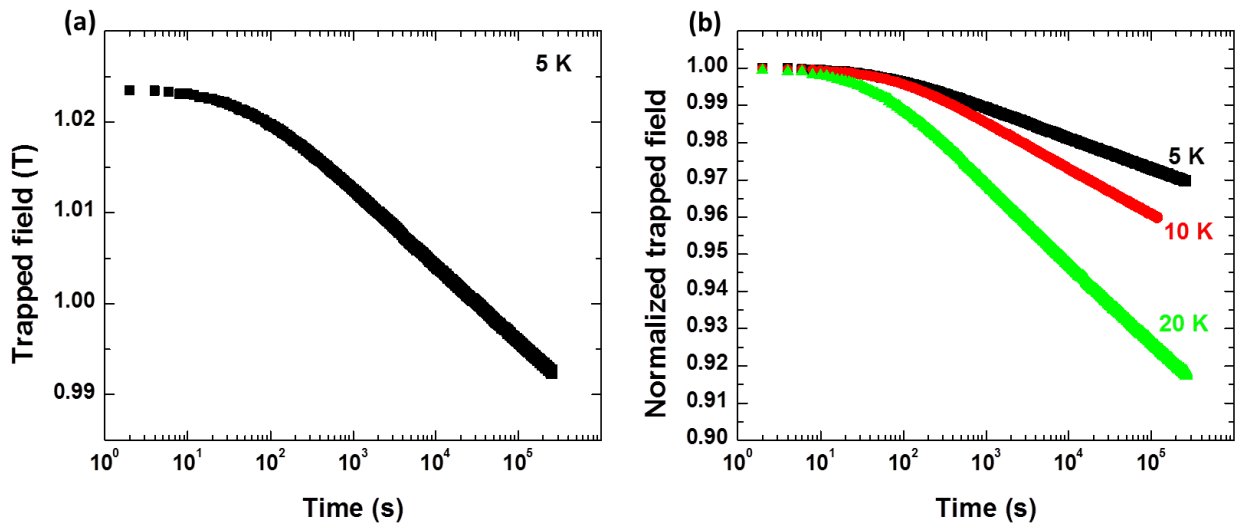

Figure 4 – Trapped magnetic field magnet creep at H2. (a) Time dependence of trapped field at 5 K and (b) normalized magnetic field creep as a function of time at 5, 10, and 20 K.

Demonstration of an iron-pnictide bulk superconducting magnet capable of trapping over 1 T

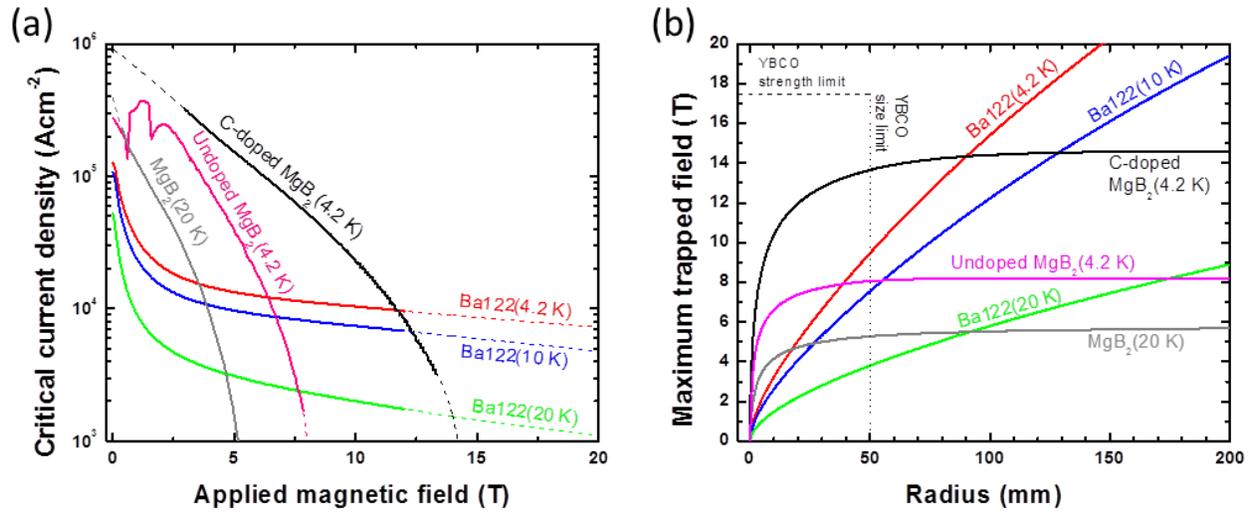

Figure 5 – Comparison between K-doped Ba122 and MgB$_2$. (a) Critical current density vs. applied magnetic field for Ba122,[10] undoped MgB$_2$,[25] and C-doped MgB$_2$ bulks.[24] Dotted lines are extrapolated data. (b) Maximum trapped field vs. radius for K-doped Ba122 and MgB$_2$ polycrystalline bulks calculated from the data in (a) for an infinite thickness cylinder.

11